\documentclass[prd,aps,twocolumn]{revtex4}
\usepackage{epsfig}
\usepackage{amsfonts}
\usepackage{rotating}
\usepackage{sparticles}
\usepackage{dcolumn}
\usepackage{bm}
\usepackage{hyperref}
\hypersetup{
    bookmarks=true,         
    unicode=false,          
    pdftoolbar=true,        
    pdfmenubar=true,        
    pdffitwindow=false,     
    pdfstartview={FitH},    
    pdftitle={My title},    
    pdfauthor={Author},     
    pdfsubject={Subject},   
    pdfcreator={Creator},   
    pdfproducer={Producer}, 
    pdfkeywords={keyword1} {key2} {key3}, 
    pdfnewwindow=true,      
    colorlinks=true,       
    linkcolor=red,          
    citecolor=cyan,        
    filecolor=magenta,      
    urlcolor=green,           
    linktocpage=true
}
\usepackage{amsmath,amssymb,amsthm,graphicx,latexsym}
\usepackage{subfigure}
\usepackage{pstricks}
\usepackage{color}

\usepackage{mathrsfs}

\newcommand{\be}{\begin{equation}}
\newcommand{\ee}{\end{equation}}
\newcommand{\bea}{\begin{eqnarray}}
\newcommand{\eea}{\end{eqnarray}}
\newcommand{\bml}{\begin{subequations}}
\newcommand{\eml}{\end{subequations}}
\newcommand{\bfig}{\begin{figure}}
\newcommand{\efig}{\end{figure}}

\newcommand{\slashed}{\hspace{-1.1ex}/}

\begin{document}

\title{Collider constraints on Gauss-Bonnet coupling in warped geometry model}
\author{{\textcolor{black}{Sayantan Choudhury}}$^{1}$, Soumya Sadhukhan$^{2}$ and {\textcolor{black}{Soumitra SenGupta}}$^{3}$}
\affiliation{$^1$Physics and Applied Mathematics Unit, Indian Statistical Institute, 203 B.T. Road, Kolkata 700 108, India}
\affiliation{$^2$Institute of Mathematical Sciences
C.I.T Campus, Taramani
Chennai, India 600113}
\affiliation{$^3$Department of Theoretical Physics,
Indian Association for the Cultivation of Science,
2A \& 2B Raja S.C. Mullick Road,
Kolkata - 700 032, India.}

\begin{abstract}
In this paper the requirement of a warp solution in an Einstein-Gauss-Bonnet 5D warped geometry is shown to fix the signature of 
Gauss-Bonnet coupling ($\alpha_{5}$). 
Further, imposing the phenomenological constraints, obtained from the 
recently observed Higgs like scalar mass as well as 
$\mu$ parameter of the decay channels ${\bf H}_{0}\rightarrow\gamma\gamma,\tau{\bar \tau}$ explored in ATLAS and CMS detectors, 
we obtain a stringent bound on $\alpha_{5}$ within $(4.8-5.1)\times 10^{-7}$.

\end{abstract}


\maketitle

The Gauss-Bonnet (GB) gravity is a good old name amongst theoretical physicists through decades, which was proposed to renormalize the 
well known Einstein's gravity at the two-loop level. For such theories the two-loop effective
 coupling physically signifies the strength of the self-interaction between the 
graviton degrees of freedom below the Ultra-Violet (UV) cut-off of the quantum theory of gravity. Such coupling
 plays a crucial role in various aspects of general relativity and particle physics in AdS space-time. Since GB correction is 
quadratic topological invariant in 4D, it will have more significant contributions in dimension $D>4$. Different proposals have been made earlier regarding the signature
 and bound of the two-loop GB coupling. 

In this work we propose a fruitful technique to 
constrain the signature and bound of 5D GB coupling in the background of the study of warped geometry model in LHC physics. Such technique can also be applicable to 
any of the higher derivative modified gravity scenarios. Here the 5D warped geometry model
have been proposed by making use of the following sets of assumptions as a building block:
\begin{itemize}
 \item The Einstein's gravity sector is modified by the introduction of quadratic Gauss-Bonnet correction.
\item  The well known ${\bf S^{1}/Z_{2}}$ orbifold compactification technique is considered.
\item We considered that the system is embedded in AdS bulk where the background warped metric has a Randall Sundrum (RS) like structure.
\item The Higgs like scalar, left handed fermion and the abelian/non-abelian gauge degrees of freedom in our model are placed in the AdS bulk 
\cite{rizzo,pomarol,okada,cheng,ssg1,ssg2}.
\end{itemize}
First we compute the warping solution to fix the signature of GB coupling. Then using the phenomenological
constraint obtained from Higgs diphoton and dilepton decay channels we also obtain a stringent 
phenomenological bound on the GB coupling which lies below the upper bound of viscosity-entropy ratio in its holographic dual version.   

To establish our proposed idea we start our discussion with the 5D action of the warped geometry two brane model given by\cite{Choudhury:2013jhep}:
\begin{widetext}
\be\begin{array}{llll}\label{eq1}
 \displaystyle S=\int d^{5}x \left[\sqrt{-g_{(5)}}\left\{\frac{M^{3}_{5}}{2}R_{(5)}+\frac{\alpha_{5}M_{5}}{2}
\left[R^{ABCD(5)}R^{(5)}_{ABCD}-4R^{AB(5)}R^{(5)}_{AB}+R^{2}_{(5)}\right]+{\cal L}^{field}_{Bulk}-2\Lambda_{5}\right\}\right.\\ \left.
\displaystyle ~~~~~~~~~~~~~~~~~~~~~~~~~~~~~~~~~~~~~~~~~~~~~~~~~~~~~~~~~~~~~~~~~~~~~~~~~~~~~~~~~~~~~~~~~~~~~~~~~+
\sum^{2}_{i=1}\sqrt{-g^{(i)}_{(5)}}\left[{\cal L}^{field}_{i}-V_{i}\right]\delta(y-y_{i})\right]
\end{array}\ee
\end{widetext}
where $i$ signifies the brane index, $i=1(\text{hidden})$, $2(\text{visible})$ and
${\cal L}^{field}_{i}$ is the Lagrangian for the fields on the ith brane with the brane tension $V_{i}$.
The background metric describing slice of the AdS is given by \cite{Choudhury:2013jhep,Lisa1,Lisa2,ssg,Kim:1999dq,Kim:2000pz},
\be\begin{array}{llllll}\label{brane}
   \displaystyle ds^{2}_{5}=g_{AB}dx^{A}dx^{B}=e^{-2A(y)}\eta_{\alpha\beta}dx^{\alpha}dx^{\beta}+r^{2}_{c}dy^{2}.
   \end{array}\ee
After solving the five dimensional {\it Einstein Gauss Bonnet} equation of motion
 at leading order in GB coupling ($\alpha_{5}$) we obtain: 
\be\begin{array}{lll}\label{eq2}
    \displaystyle A(y)=k_{\alpha}r_{c}|y|=\sqrt{\frac{3M^{2}_{5}}{16\alpha_{5}}\left[1-\left(1+\frac{4\alpha_5 \Lambda_5}{9M^{5}_{5}}\right)^{\frac{1}{2}}\right]}r_{c}|y|.
   \end{array}\ee
In the limit $\alpha_{5}\rightarrow 0$, we retrieve the same result as in the case of RS model with
$k_{\alpha}\rightarrow k_{RS}=\sqrt{-\frac{\Lambda_{5}}{24M^{3}_{5}}}$ \cite{Lisa1,Lisa2}. 
First we observe from Eq~(\ref{eq2}) that the warping solution for AdS space-time with $\Lambda_{5}<0$ requires the signature of the GB coupling
$\alpha_{5}$ to be positive.

 Next we focus our attention to the holographic dual boundary Conformal Field Theory (CFT) of the
present bulk AdS theory where the viscosity-entropy ratio can be calculated in presence of GB coupling using the well known 
{\it Kubo formula} as \cite{Choudhury:2013,Myers:2008PRD}:
\be\begin{array}{llll}\label{eq4}
    \displaystyle \frac{\eta}{S}=\frac{1}{4\pi}\left(1-4\alpha_{5}\right)+{\cal O}(\alpha^{2}_{5}).
   \end{array}\ee
In the limit $\alpha_{5}\rightarrow 0$ Eq~(\ref{eq4})
reduces to the usual result in Einstein's gravity, $\frac{\eta}{S}\rightarrow\frac{1}{4\pi}$ \cite{Son1:2001,Buchel:2004,Son2:2005}. As GB coupling comes from two-loop perturbative correction to the Einstein's gravity
the numerical value of $\alpha_{5}$ should be small. This also implies ${\cal O}(\alpha^{2}_{5})$ gives sufficiently small 
sub-leading contribution to the Eq~(\ref{eq4}). Moreover, $\frac{\eta}{S}>0$ criteria 
constraints the upper bound of GB coupling $\alpha_{5}$ to be less than $1/4$. 

We now explored the possible upper \& lower bound of the GB coupling obtained from the phenomenological 
constraints in the ATLAS \cite{ATLASMa} \& CMS \cite{CMSAp} data from the recent run of LHC.
 For this we perform the dimensional reduction of the 5D fields using the the well known KK decomposition technique
 to extract the SM fields in the effective 4D theory.

\begin{figure}[t]
\centering
\includegraphics[width=8.5cm,height=6cm]{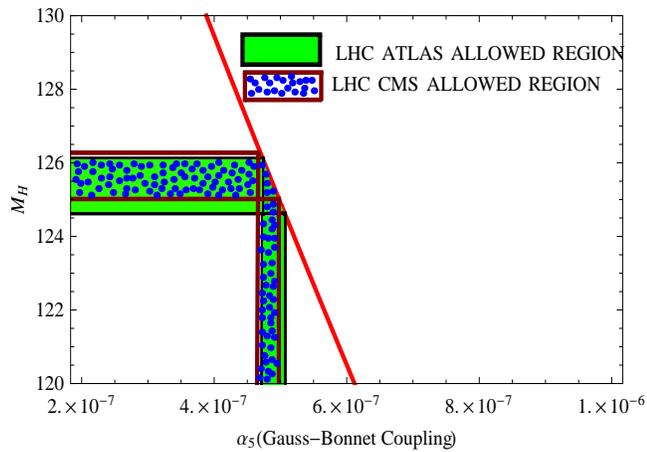}
\caption{Variation of zeroth mode KK mass of bulk scalar field with respect to Gauss-Bonnet coupling ($\alpha_{5}$ ). The green shaded
and the blue dotted region shows the LHC ATLAS and CMS allowed region of recently observed Higgs at 125 GeV respectively. Here we use all the
statistical and systematic errors of the corresponding detectors of LHC for the numerical estimations. For numerical estimations throughout the paper we fix
the compactification radius $r_{c} \approx 1$ in the Planck unit.
}
\label{f1}
\end{figure}

To proceed further we assume that the spontaneous symmetry breaking mechanism is taking place in 5D by which, the bulk
scalar, bulk fermions and gauge bosons are getting mass. Such scenarios have been elaborately 
discussed in various earlier works \cite{rizzo,pomarol,okada,cheng,ssg1,ssg2}. This scenario
has the advantage of accommodating positive brane tension of the visible brane and thus ensures stability
of the proposed model \cite{ssg2,ssg3}. 
 
The mass generation procedure can be
 described from action of the scalar $\xi$ and its coupling with a fermion $\eta$ as:
\begin{align}\label{b1}
   \displaystyle S_{\xi} \supset \int d^{5}x\sqrt{-g_{(5)}}\left\{\frac{1}{2} \left[g^{AB} D_{A}{\xi}(x,y)
D_{B}{\xi}(x,y)
\nonumber\right.\right.\\ \left.\left.\displaystyle-m^{2}_{\xi}{\xi}^{2}(x,y)\right] + \lambda_{\xi}{\xi}^{4}(x,y) +  y_{\eta} \xi \eta \bar{\eta} \right\}
\end{align}
where the covariant derivative is defined as:
\begin{align}\label{covder}
D_{A} \equiv \partial_{A} + i g_{1}^{\prime} T^{b} W^{b}_{A} + i g_{2}^{\prime} Y B_{A}
\end{align} which is the suitable linear combination of W and B vector fields give rise to
 $W^{+},W^{-},Z$ and photon fields in 5D. Additionally, in Eq~(\ref{covder})
 $g_{1}^{\prime}$ and $g_{2}^{\prime}$ signifies the unbroken ${\cal SU}(2)$ and ${\cal U}(1)$
 gauge couplings while $ T^{b}$ and $ Y$ are the corresponding generators of the gauge groups associated with the W and B vector boson fields.
 Here $m^{2}_{\xi}$ signifies the mass parameter of the field $\xi$ before spontaneous symmetry breaking, $\lambda_{\xi}$
 be the self-interaction strength of scalar ($\xi$) and $y_{\eta}$ represents the Yukawa coupling.
Now after symmetry breaking mechanism by writing, $ \xi= v_1 +\bf H $ and $ \eta=  \Psi$
the 5D action stated in Eq~(\ref{b1}) can be recast as:
\begin{align}\label{b2}
   \displaystyle  S_{\bf SSB} \supset \int d^{5}x\sqrt{-g_{(5)}}\left\{\frac{1}{2} \left[g^{AB} \partial_{A}{\bf H}(x,y)
\partial_{B}{\bf H}(x,y)
\nonumber\right.\right.\\ \left.\left.\displaystyle+m^{2}_{h}{\bf H}^{2}(x,y)\right]  +  m_f \Psi \bar{\Psi} + y_{\eta}{\bf H }\Psi \bar{\Psi}  + h_{w}{\bf H}W^{+}W^{-}\right\}
\end{align}
Most importantly, in the effective 4D counterpart the photon field has to be massless and also it appears that for such a 
physical prescription, the 5D mass parameter must be zero. Here $v_1$ is the VEV of $\xi$ field. $h_{w}$ which characterizes
the coupling strength of the ${\bf H}W^{+}W^{-}$ interaction and the physical mass parameter for the scalar and fermionic degrees of freedom after spontaneous symmetry breaking can be written as:
\begin{align}\label{b3}
\displaystyle  m^{2}_{h}= -\frac{m^{2}_{\xi}}{2 \lambda_{\xi}},~~~
\displaystyle m_f=y_{\eta} v_1,~~~
\displaystyle h_{w}= \frac{1}{2}{g_{1}^{\prime}}^2 v_1.
\end{align}

Further the recent observation of a resonance named Higgs at the LHC \cite{ATLAS07,CMS03} with mass around 125 GeV will help us to compare the new particle 
to the zeroth mode of the KK mass spectrum of scalar degrees of freedom (${\bf H}_{0}$) in presence of GB coupling. 
This zeroth mode scalar mass obtained as \cite{Choudhury:2013jhep}: 
\be\begin{array}{llll}\label{massasdphi}
   \displaystyle  m_{s}=M_{H}\approx \left(\frac{1}{2}\sqrt{4+\frac{m_h^2}{k^2_{\alpha}}}
-\frac{3}{4}\right)\pi k_{\alpha} e^{-k_{\alpha} r_{c}\pi}.
   \end{array}\ee
In Fig~(\ref{f1}) we have demonstrated the behaviour of 
zeroth mode Higgs like scalar mass appearing in the 4D effective version of the warped geometry model with respect to the GB coupling by the red curve.
We have also shown the present status of our proposed model after applying the Higgs mass constraint $M^{CMS}_{H}=(125.7\pm 0.3)^{+0.3}_{-0.3}$~GeV \cite{CMSAp}
 and $M^{ATLAS}_{H}=(125.5 \pm 0.2)^{ +0.5}_{-0.6}$~GeV \cite{ATLASMa}
obtained from ATLAS \& CMS data by {\it green shaded} and {\it blue dotted} region respectively. This allows to put a stringent bound on the GB coupling within
$4.6\times 10^{-7}<\alpha_{5}<5.1\times 10^{-7}$ as shown in a very tiny region in Fig~(\ref{f1}). Additionally, this also satisfies the constraint on GB coupling   
obtained from Eq~(\ref{eq4}).

Next we apply the constraint on GB coupling from different decay channels of Higgs where 
interactions  between various KK zero modes play a crucial role. The effective field theory provides such interactions below UV cut-off.
Most importantly at the tree level the scalar KK zero mode is not coupled to 
any of the gauge boson. However at loop level such coupling indeed appears signifying some possible non-trivial consequences.
 Additionally, it appears that 
the total decay width is dominated by the contribution of two fermion decay mode at the tree level.
For computing the higgs decay to two photons and two gluons the contribution of top quark triangle
 loop diagram as well as decay into two W boson channels are considered. We also assume that the effects from the other fermionic degrees of freedom in the loop level analysis are sufficiently small
because the other quarks and leptons being much lighter will contribute much less in the decay width and branching ratio.
 Here we use a generic ansatz for KK decomposition for the bulk 5D fields as,
\be\begin{array}{lllll}\label{KK41bs}
\displaystyle {\bf K}_{\beta}(x,y)=\sum^{\infty}_{n=0}{\bf K}^{(n)}_{\beta}(x)~\frac{\chi^{(n)}_{{\bf K}_{\beta}}(y)}{\sqrt{r_{c}}}. 
   \end{array}\ee 
where $\beta= \ 1 (\text{Higgs like scalar}), \ 2 (\text{Left handed fermion}),\\ \ 3 (\text{Abelian gauge field}), \ 4 (\text{Non-abelian gauge field})$.

\begin{figure}[t]
\centering
\includegraphics[width=8.5cm,height=6cm]{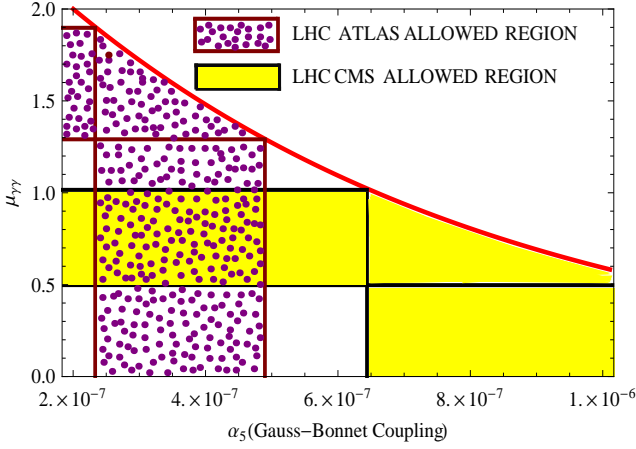}
\caption{Variation of $\mu_{\gamma\gamma}$ for $H\rightarrow\gamma\gamma$ 
 with respect to Gauss-Bonnet coupling ($\alpha_{5}$) for AdS space
($\Lambda_{5}<0$). The yellow shaded region shows the LHC CMS allowed region. The violet dotted region represents the LHC ATLAS
allowed region for our proposed model.
}
\label{f2}
\end{figure}


\begin{figure}[t]
\centering
\includegraphics[width=8.5cm,height=6cm]{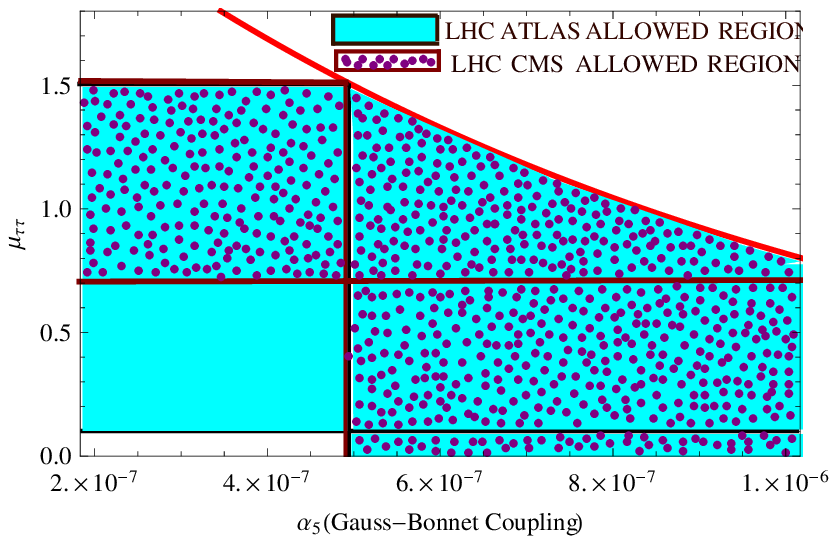}
\caption{Variation of $\mu_{\gamma\gamma}$ for $H\rightarrow\tau{\bar \tau}$ with respect to Gauss-Bonnet coupling ($\alpha_{5}$) for AdS space
($\Lambda_{5}<0$). The aqua shaded region shows the LHC ATLAS allowed region. The violet dotted region represents the LHC CMS
allowed region for our proposed model. 
}
\label{f3}
\end{figure}

In the present context to find the Yukawa interaction between two KK zero mode fermions and one scalar degrees of freedom,
the vertex factor of the 4D counterpart of the 5D action after KK decomposition can be obtained from: 
\begin{align}\label{eqb}
{\cal Q}^{000}:= \int^{+\pi}_{-\pi}dy~ e^{4A(y)}\chi^{(0)}_{{\bf K}_{1}}(y)\chi^{(0)\star}_{{\bf K}_{2}}(y)\chi^{(0)}_{{\bf K}_{2}}(y).  
\end{align}
Similarly the interaction between two KK zero mode fermions and one massless abelian/non-abelian gauge boson can be obtained from 
\be\begin{array}{lllll}\label{ggcuyt}
    \displaystyle {\cal G}^{000}:=\int^{+\pi}_{-\pi}dy~ e^{A(y)}\chi^{(0)\star}_{{\bf K}_{2}}(y)\chi^{(0)}_{{\bf K}_{3/4}}(y)\chi^{(0)}_{{\bf K}_{2}}(y).
   \end{array}\ee
Also the interaction of one KK zeroth mode scalar and two KK zero mode massive non-abelian gauge bosons can be obtained from:
 \be\begin{array}{lllll}\label{ggcuyt}
    \displaystyle {\cal W}^{000}:=\int^{+\pi}_{-\pi}dy~ \chi^{(0)}_{{\bf K}_{1}}(y)\chi^{(0)}_{{\bf K}_{4}}(y)\chi^{(0)}_{{\bf K}_{4}}(y) 
   \end{array}\ee and the interaction strength of a vertex of one photon zero mode with 2 massive weak boson zero modes can be expressed as: 
\be\begin{array}{lllll}\label{ggcuyt}
    \displaystyle {\cal A}^{000}:=\int^{+\pi}_{-\pi}dy~ \chi^{(0)}_{{\bf K}_{3}}(y)\chi^{(0)}_{{\bf K}_{4}}(y)\chi^{(0)}_{{\bf K}_{4}}(y).
   \end{array}\ee
In the interaction integrals for KK zero mode as stated in Eq~(\ref{eqb}) \& Eq~(\ref{ggcuyt})
we use the following expressions for the KK zero mode wave functions for the different field contents:
\be\label{normala}
\chi^{(0)}_{{\bf K}_{\beta}}(y)=
\left\{
	\begin{array}{ll}
                    \displaystyle N_s & \mbox{ \it for $\beta=1$}  \\
         \displaystyle N_{f}e^{-\nu A(y)} & ~\mbox{\it for $\beta=2$} \\
         \displaystyle  \frac{1}{\sqrt{2\pi}} & ~\mbox{\it for $\beta=3,4$}.
          \end{array}
\right.
\ee
where $N_f=\sqrt{\frac{2\left[e^{(1- 2\nu)k_{\alpha} r_{c}\pi}-1
\right]}{\left(1 - 2\nu \right)k_{\alpha} r_{c}}} $ and $N_s=\sqrt{\frac{k_{\alpha} r_{c}}{1-e^{-2k_{\alpha} r_{c}\pi}}} $.
Also we use $\nu=\frac{m_f}{k_{\alpha}}$, where $m_f$ is defined in Eq~(\ref{b3}). 
Finally the effective 4D coupling for the interactions stated in Eq~(\ref{eqb}) \& Eq~(\ref{ggcuyt}) can be written as: 
\bea\label{fq}
F_{\cal Q} &=& y_{\eta}{\cal Q}^{000}=\frac{2 N_{f}^2 N_{s}}{m_f r_c} \sinh{(m_f r_c \pi)},\\
F_{\cal G}&=&\frac{g_{f}{\cal G}^{000}}{\sqrt{r_c}}=\frac{g_{f}N_{f}^{2}}{\sqrt{2 \pi r_c}} \frac{\sinh{(k_{\alpha}-2 m_f)\pi r_c}}{(k_{\alpha}-2 m_f) r_c},\\
F_{\cal W}&=&h_{w}{\cal W}^{000}=h_{w}N_{s},\\
F_{\cal A}&=&\frac{g_e}{r_c^{\frac{3}{2}}}{\cal A}^{000} =  \frac{g_e}{\sqrt{2 \pi r_c^3}}
\eea 
where $y_{\eta}$ is the 5D Yukawa coupling, $g_f$ is the gauge coupling
 which is different for different gauge groups and $g_e$ is the 5D gauge coupling of spontaneously broken ${\cal U}(1)$.

So obtaining the couplings of the zeroth mode massive scalar boson
 with other particles we calculate the
 decay width and the branching ratio in various channels. We can
 also obtain the production cross section of this particle, which can
 be the resonances found at 125 GeV describing
 the SM Higgs boson. Using these results together we can 
 get the expression for the LHC observable $\mu$ parameter (described in the Appendix)
 for di-photon and dilepton channels. 
Hence comparing the derived $\mu$ parameter with the experimental results obtained from the CMS and ATLAS 
 one can get further stringent constraints on GB coupling for suitable choice of other parameters.

The Higgs like scalar candidate in our proposed model couple to the KK zeroth mode of other massive bulk
 fermions and massive gauge bosons at the tree level. Different SM fermion masses can be obtained from different 5D bulk fermion mass 
parameter. Using these inputs we can compute the decay width of the dacay channel of the scalar KK zero mode to various fermion KK zero modes using the vertex function
 explicitly mentioned in Eq.~(\ref{fq}). Using Eq~(\ref{b3},\ref{fq}), the decay width to ith fermionic channel can be written as
\begin{align}
\Gamma({\bf H}_{0} \rightarrow f_i \bar{f_i})= \frac{N_c}{8 \pi} F^{2}_{{\cal Q}_{i}}m_{s}\left(1-\frac{4 m_{f_{i}}^{2}}{m_{s}^{2}}\right)^{\frac{3}{2}}
\end{align}
where $i=1,2,3,4$ correspond to the tau lepton, bottom, charm and top quark, analogous to SM fermions, respectively and also $m_{f_{i}}$ represent the different
fermion masses of the ``i''th species. $N_c$ is 3 for all quarks and 1 for leptons. 
 Summimg over the all leading contributions total fermionic decay width can be written as:
\begin{align}
\Gamma_{\text{fer}} \approx \sum^{4}_{i=1}\Gamma({\bf H}_{0} \rightarrow f_i \bar{f_i})= \frac{N_c m_s}{8 \pi}
 \sum^{4}_{i=1} F^{2}_{{\cal Q}_{i}}\left(1-\frac{4 m_{f_{i}}^{2}}{m_{s}^{2}}\right)^{\frac{3}{2}}.
\end{align}
Another considerable contribution comes from the scalar decay to the zeroth mode massive $W^{+},W^{-}$ bosons.
 Here the Higgs like scalar mostly decays to one on-shell ($W^{+}$) and another off-shell ($W^{-\ast}$) W bosons.
 This off shell W further decays to fermion pairs. This is expressed as:
\begin{align}
\Gamma({\bf H}_{0} \rightarrow W^{+} W^{-\ast})= \frac{m_s^4}{48 m_W^2} F_{\cal W}^2 F_{\cal G}^2 T\left(m_{W}^2,m_s^2\right)
\end{align}
where the explicit form of $T(m_{W}^2,m_s^2)$ is given by:
\begin{widetext}
\be\begin{array}{llll}\label{fh2}
 \displaystyle T\left(m_{W}^2,m_s^2\right)= \int d(k^{2})
 \frac{[\lambda^{\frac{3}{2}}(m_W^2, k^2,m_s^2) + \lambda^{\frac{1}{2}}(m_W^2, k^2,m_s^2) \frac{12 k^2 m_W^2}{m_s^4}]}{(k^2-m_W^2)^2 + {\Gamma}_W^2 m_W^2}
\end{array}\ee 
\end{widetext}
with a new parameter
\be\lambda(m_W^2, k^2,m_s^2)= \left(1-\frac{m_W^2}{m_s^2}-\frac{k^2}{m_s^2}\right)^2 - \frac{4 m_W^2 k^2 }{m_s^4}.\ee
Additionally, $m_{W}$ and $\Gamma_W$ characterizes the mass and decay width respectively corresponding to the KK zero mode W boson.
Consequently the total decay width of this scalar can be written as:
\begin{align}
\Gamma_{\text{total}} \approx \Gamma({\bf H}_{0} \rightarrow W^{+} W^{-\ast}) + \Gamma_{\text{fer}}.
\end{align}
It is important to mention here that in the present context the 
scalar decay in to two photons loop diagrams of both zeroth mode W boson and top quark will contributing.
 However the contribution from the W loops are found to be dominant. Taking into account only the 
dominant contribution from the W in computing the decay width we get:
 \begin{align}
\Gamma({\bf H}_{0} \rightarrow \gamma \gamma) = \frac{m_{s}^3}{3456 {\pi}^5 m_{W}^{4}} F^{2}_{\cal A}  F^{2}_{\cal W} \left[U\left(\frac{4 m_{W}^2}{m_s^2}\right)\right]^2
\end{align}
 where the functional form of $U(q)$ is given by:
\begin{align}\label{fh2}
 U\left(q\right)=\left[2 + 3 q + 3 q \left(2 - q\right)\left\{sin^{-1}\frac{1}{\sqrt{q}}\right\}^2\right].
\end{align}
 with $q=\frac{4 m_{W}^2}{m_s^2} > 1 $. Further, to calculate the decay width of the zeroth mode scalar to two
 massless gluons which are zeroth KK modes of the
 non abelian fields with strong gauge interaction
 strength, the vertex factor of the two fermion zero mode (mainly top like one) and one gauge boson
 zero mode is to be taken into account. Consequently the digluon decay width can be computed as: 
\begin{align}\label{fh1}
\Gamma({\bf H}_{0} \rightarrow gg) = \frac{m_{f_4}^2 N_{g}}{256 \pi^{5} m_s}
 F_{{\cal Q}_{4}}^2 F^{4}_{{\cal G}_{g}} \left[ D\left(\frac{m_{f_4}^2}{m_s^2}\right)\right]^2
\end{align}
where $N_{g}$ is the colour factor which is 8 for gluon channel. In this context, we define a new
function:
\begin{align}\label{fh2}
 D\left(z\right)=\left[1+ \left(1-4z\right)\left\{sin^{-1}\frac{1}{2\sqrt{z}}\right\}^2\right]
\end{align}
with $z=\frac{m_{f_4}^2}{m_s^2} > 1/4$.
 In LHC gluon-gluon fusion is the dominant mechanism for the scalar production. 
Differential production cross section of the scalar considered here in a hadronic collider like LHC  is given as \cite{hh},
\begin{align}\label{diff}
\displaystyle \frac{d\sigma}{dy}(p\bar{p}\rightarrow {\bf H}_{0} +X)= \frac{{\pi}^2\Gamma[{\bf H}_{0} \rightarrow \text{gg}]}
{8m_s^3} g_p(x_p,m_s^2)g_{\bar{p}}(x_{\bar{p}},m_s^2)
\end{align}
where we define gluon momentum fraction as:
\begin{align}
\displaystyle x_p= \frac{m_s e^y}{\sqrt{s}},~~~~ x_{\bar{p}} = \frac{m_s e^{-y}}{\sqrt{s}}.
\end{align}
In Eq~(\ref{diff}) $y$ represents the rapidity of the scalar,
 $ g_p(x_p,M_s^2)$ is the gluon distribution function in proton evaluated at the gluon momentum fraction $x_p$
 and $\sqrt{s}$ signifies the beam energy. 
Assuming the rapidity to be same for both the scalar
and the SM Higgs, the ratio of the production cross section in the $\mu$ parameter can be
 written only as the ratio of the gluon-gluon decay width of this scalar
 to the Higgs width in that channel. 
Substituting all of these inputs in Eq~(\ref{mup}) (Please see the appendix), we can write the explicit expression for the LHC observable $\mu$ parameter.

Implication of the recent LHC results in diphoton and dilepton decay channels
 of the Higgs like scalar in this model are explicitly shown in  
Fig.~(\ref{f2}) and Fig.~(\ref{f3}) respectively. The allowed regions which are obtained
 for the 5D GB coupling ($\alpha_{5}$), in these representative figures
 are also consistent with the phenomenological bound of $\alpha_{5}$ that we have obtained from the
 Higgs mass constraints. For the ${\bf H}_{0}\rightarrow\gamma \gamma $ decay channel in ATLAS,
 a region with $2.3\times 10^{-7}<\alpha_{5}<4.9\times 10^{-7}$ is
 allowed and this region contains the region that we have already obtained from the
 Higgs mass constraint. But CMS shows a slightly shifted region of
 $\alpha_{5}>6.4\times 10^{-7}$ than the allowed one. As CMS and ATLAS
 results differ a bit in this particular channel, no strict constraint can be found to determine the allowed region.
 With more data coming in the next run of the LHC we can pinpoint the allowed region for $\alpha_{5}$
 in the diphoton channel. For the ${\bf H}_{0}\rightarrow\tau {\bar\tau}$ decay channel the experimental
 upper limits on the observed $\mu$ parameter value both in CMS and ATLAS allows the region with
$\alpha_{5}>4.8\times 10^{-7}$. This region contains some part of
 the region allowed by the diphoton channel data and Higgs mass. But 
in this channel experimental ranges for the $\mu$ parameter are much broader indicating lesser 
statistical accuracy. This clearly implies that lower limits on the $\mu$ parameter are unable
 to constraint the upper bound of $\alpha_{5}$ in ${\bf H}_{0}\rightarrow\tau {\bar\tau}$ channel. It is expected
 that more data will help to give stronger upper bound of $\alpha_{5}$ in
 this channel. Now combining the constraints from all the three cases,namely, $\mu$
 parameter values in the ${\bf H}_{0}\rightarrow\gamma \gamma$ and ${\bf H}_{0}\rightarrow\tau {\bar\tau}$ decay channels and mass of the
 resonance discovered near 125 GeV we can constrain the allowed region of $\alpha_{5}$
 within $4.8\times 10^{-7}<\alpha_{5}<5.1\times 10^{-7}$. Additionally, this bound
 also satisfies the criterion obtained from viscosity-entropy ratio as mentioned in Eq~(\ref{eq4}).

To summarize, we say that the perturbative higher order gravity correction to Einstein's gravity can also be examined
 through collider experimental tests. Using the tools mentioned in this paper one can directly check the validity of a higher order gravity 
or any modified gravity model and also constrain the couplings associated with such higher order gravity corrections. 
Thus, in this work, by applying the requirements for the warping solution of the metric, we have explicitly  
shown that for 
Gauss-Bonnet (GB) gravity, the associated two-loop coupling is always positive and less than $1/4$.
Further, imposing the phenomenological constraint from Higss dilepton and diphoton decay channels and also from possible Higgs mass constraint at 125 GeV
we obtain a stringent bound on the GB coupling which satisfies the above criteria as well. 
This result is consistent with similar bounds obtained from solar system constraint \cite{ssg4}. This analysis therefore 
determines the signature of GB coupling and brings out the phenomenological constraint on the value of this parameter in the context of recent LHC experiment.


\section*{Acknowledgments}
SC thanks Council of Scientific and
Industrial Research, India for financial support through Senior
Research Fellowship (Grant No. 09/093(0132)/2010). SS thanks Institute of Mathematical Sciences for Senior Research Fellowship. 
We also acknowledge illuminating discussions with Dilip Kumar Ghosh.

\section*{Appendix}
 The $\mu$ parameter is designed for comparison of the number of events obtained from a decaying particle to that of the SM expectation.
 For more than one decay channels present branching ratio (BR) of a particle, X, to any channel $l_1 l_2$ is defined as:
\begin{align}
BR(X \rightarrow l_1 l_2)=\frac{ \Gamma (X \rightarrow l_1 l_2)}{\Gamma(X \rightarrow \text{all decay channels})}
\end{align}
where 
$\Gamma (X \rightarrow l_1 l_2) = \text{Decay width of}~ (X \rightarrow l_1 l_2)$.
Now total production cross section of a particle multiplied by its branching ratio to a decay channel and the luminosity present in the experiment gives us the number of decay products from the signal.
 So for the diphoton channel in Higgs search the $\mu$ parameter of the model can be expressed as: 
\be\label{mup}
\mu=\frac{\Sigma ({\bf pp}\rightarrow {\bf H}_{0})}{\Sigma({\bf pp}\rightarrow {\bf H_{SM}})}
 \times\frac{BR({\bf H}_{0} \rightarrow \gamma \gamma)}{BR(\bf H_{SM} \rightarrow \gamma \gamma)}
 \ee for fixed collider luminosity. Here $\Sigma$ represents the production cross section, ${\bf H_{SM}}$ signifies the SM Higgs and ${\bf H}_{0}$ is used for zeroth KK mode 
Higgs like particle in the warped geometry model proposed in Eq~(\ref{eq1}). Similar expression appears for the dilepton channel.



\end{document}